\documentclass[epsf,12pt]{article}

\usepackage{amssymb}

\usepackage[dvips]{graphicx}

\begin{document}

\title
{Two-loop Gell-Mann--Low function of N=1 supersymmetric Yang-Mills
theory, regularized by higher covariant derivatives.}

\author{A.B.Pimenov \thanks{E-mail:$pimenov.phys.msu@mail.ru$},
K.V.Stepanyantz\thanks{E-mail:$stepan@phys.msu.ru$}}

\maketitle

\begin{center}
{\em Moscow State University, physical faculty,\\
department of theoretical physics.\\
$119992$, Moscow, Russia}
\end{center}

\begin{abstract}
Two-loop Gell-Mann--Low function is calculated for N=1
supersymmetric Yang--Mills theory, regularized by higher covariant
derivatives. The integrals, which define it, are shown to be
reduced to total derivatives and can be easily calculated
analytically.
\end{abstract}


\section{Introduction.}
\hspace{\parindent}

It is well known that supersymmetry essentially improves the
ultraviolet behavior of a theory. For example, even in theories
with unextended supersymmetry, it is possible to propose the form
of the $\beta$-function exactly to all orders of the perturbation
theory. This proposal was first made in Ref. \cite{NSVZ_Instanton}
as a result of investigating instanton contributions structure.
For the $N=1$ supersymmetric Yang-Mills theory, this
$\beta$-function, called the exact
Novikov--Shifman--Vainshtein--Zakharov (NSVZ) $\beta$-function is

\begin{equation}\label{NSVZ_Beta}
\beta(\alpha) = - \frac{3 C_2 \alpha^2 }{2\pi(1- C_2\alpha/2\pi)}.
\end{equation}

\noindent With the dimensional reduction, this expression
coincides with the result of explicit calculating the function

\begin{equation}\label{Beta_Definition}
b(\alpha) = \frac{d\alpha}{d\ln\mu},
\end{equation}

\noindent up to the four-loop approximation
\cite{ThreeLoop1,ThreeLoop2,ThreeLoop3,JackJones}, if a special
choice of the renormalization prescription is used. Here $\alpha$
is a renormalized coupling constant, and $\mu$ is a normalization
point. However, function (\ref{Beta_Definition}) is scheme
dependent. The physical $\beta$-function is obtained from it only
if the generating functional does not depend on the normalization
point, and some special boundary conditions, which involve knowing
finite parts of Green functions, are imposed. However, as a rule,
the divergent part of the effective action in $\overline{MS}$
scheme is only calculated with the dimensional reduction. We also
note that the two-loop $\beta$-function was also calculated with
the differential renormalization \cite{Mas}.

Investigation of the $N=1$ supersymmetric electrodynamics (up to
the four-loop approximation) \cite{hep,tmf2,ThreeLoop,Pimenov}
shows that the exact NSVZ $\beta$-function coincides with the
Gell-Mann-Low function.

This function is scheme independent, and can be calculated with an
arbitrary regularization. However, using the higher derivative
regularization \cite{Slavnov,Slavnov_Book} is the most convenient.
That matter is that with the higher derivative regularization all
integrands, appearing in calculating the Gell-Mann--Low function,
are total derivatives. It was first noted in Ref.
\cite{ThreeLoop}. Therefore, in fact, these integrals can be
easily taken. In the electrodynamics this can be partially
explained by a method, based on substituting solutions of Ward or
Slavnov--Taylor identities into the Schwinger--Dyson equations
\cite{SD}. However, for a complete proof it is necessary to
propose existence of a new identity for the Green functions, which
origin is so far unclear \cite{Identity}. A similar identity
\cite{SDYM} also appears in calculating a contribution of the
matter superfields in a non-Abelian theory. It exists because
integrals, defining the two-point Green function are reduced to
total derivatives. Nevertheless, this statement was not yet
verified in the non-Abelian case. And, in particular, the two-loop
calculation of the Gell-Mann--Low function with the higher
covariant derivatives (which is made in this paper) allows
elucidating whether similar fact takes place in the non-Abelian
case.

We note that using the higher covariant derivative regularization
in non-Abelian theories is technically complicated. That is why
such a regularization was applied only once, for the one-loop
calculation in the (non-supersymmetric) Yang--Mills theory
\cite{Ruiz}. Taking into account comments, made in subsequent
papers \cite{Asorey,Bakeyev,PhysLett}, the result of the
calculation coincided with the the standard expression for the
one-loop $\beta$-function (although in original paper \cite{Ruiz}
the authors affirm that it is not so). As we already mentioned, a
purpose of this paper is calculating the Gell-Mann--Low function
for the $N=1$ supersymmetric Yang--Mills theory with the higher
derivative regularization in the two-loop approximation. This
function is defined as follows. Due to the Slavnov--Taylor
identities a contribution to the renormalized effective action,
corresponding to the two-point Green function of the gauge field,
obtained using the background field method, is

\begin{equation}\label{D_Definition}
-\frac{1}{8\pi}\mbox{tr} \int d^4\theta\,\frac{d^4p}{(2\pi)^4}
{\bf V}(-p)\,\partial^2 \Pi_{1/2} {\bf
V}(p)\,d^{-1}(\alpha,\mu/p),
\end{equation}

\noindent where ${\bf V}$ is the background field, and
$\partial^2\Pi_{1/2}$ is a supersymmetric transverse projection
operator. Then the Gell-Mann--Low function is defined by

\begin{equation}\label{Gell-Mann-Low_Definition}
\beta\Big(d(\alpha,\mu/p)\Big) = \frac{\partial}{\partial \ln p}
d(\alpha,\mu/p).
\end{equation}

This paper is organized as follows.

In Sec. \ref{Section_SUSY_QED} we recall basic information about
the $N=1$ supersymmetric Yang-Mills theory, the background field
method, and the higher derivatives regularization. Calculating a
two-loop contribution to the Gell-Mann--Low function is described
in Sec. \ref{Section_2Loop}. Diagrams with counterterms insertions
are calculated in Sec. \ref{Section_Anomaly_Puzzle} exactly to all
orders of the perturbation theory. A brief discussion of the
results is given in the conclusion.


\section{$N=1$ supersymmetric Yang-Mills theory, background field method
and higher derivative regularization} \label{Section_SUSY_QED}
\hspace{\parindent}

The $N=1$ supersymmetric Yang-Mills theory in the superspace is
described by the action

\begin{eqnarray}\label{SYM_Action}
&& S = \frac{1}{2 e^2} \mbox{Re}\,\mbox{tr}\int
d^4x\,d^2\theta\,W_a C^{ab} W_b.
\end{eqnarray}

\noindent Here the superfield $W_a$ is a supersymmetric analogue
of the gauge field stress tensor. It is defined by

\begin{equation}
W_a = \frac{1}{32} \bar D (1-\gamma_5) D\Big[e^{-2V}
(1+\gamma_5)D_a e^{2V}\Big],
\end{equation}

\noindent where $V$ is a real scalar superfield, which contains
the gauge field $A_\mu$ as a component, and

\begin{equation}
D = \frac{\partial}{\partial\bar\theta} -
i\gamma^\mu\theta\,\partial_\mu
\end{equation}

\noindent is a supersymmetric covariant derivative. In our
notation, the gauge superfield $V$ is decomposed with respect to
the generators of a gauge group $T^a$ as $V = e\, V^a T^a$, where
$e$ is a coupling constant. The generators are normalized by the
condition

\begin{equation}
\mbox{tr}(t^a t^b) = \frac{1}{2} \delta^{ab}.
\end{equation}

Action (\ref{SYM_Action}) is invariant under the gauge
transformations

\begin{equation}
e^{2V} \to e^{i\Lambda^+} e^{2V} e^{-i\Lambda},
\end{equation}

\noindent where $\Lambda$ is a chiral superfield.

For quantization of this model it is convenient to use the
background field method. The matter is that the background field
method allows calculating the effective action without manifest
breaking of the gauge invariance. In the supersymmetric case it
can be formulated as follows \cite{West, Superspace}: Let us make
a substitution

\begin{equation}\label{Substitution}
e^{2V} \to e^{2V'} \equiv e^{\mbox{\boldmath${\scriptstyle
\Omega}$}^+} e^{2V} e^{\mbox{\boldmath${\scriptstyle \Omega}$}},
\end{equation}

\noindent in action (\ref{SYM_Action}), where
$\mbox{\boldmath${\Omega}$}$ is a background scalar superfield.
Expression for $V'$ is a complicated nonlinear function of $V$,
$\mbox{\boldmath$\Omega$}$, and $\mbox{\boldmath$\Omega^+$}$. We
do not interested in explicit form of this function:

\begin{equation}
V' = V'[V,\mbox{\boldmath$\Omega$}].
\end{equation}

\noindent (For brevity of notation we will not explicitly write
the dependence on $\mbox{\boldmath$\Omega$}^+$ here and below.)
The obtained theory will be invariant under the background gauge
transformations

\begin{equation}\label{Background_Transformations}
V \to e^{iK} V e^{-iK}; \qquad e^{\mbox{\boldmath${\scriptstyle
\Omega}$}} \to e^{iK} e^{\mbox{\boldmath${\scriptstyle \Omega}$}}
e^{-i\Lambda};\qquad e^{\mbox{\boldmath${\scriptstyle \Omega}$}^+}
\to e^{i\Lambda^+} e^{\mbox{\boldmath${\scriptstyle \Omega}$}^+}
e^{-iK},
\end{equation}

\noindent where $K$ is a real superfield, and $\Lambda$ is a
chiral superfield.

Let us construct the background chiral covariant derivatives

\begin{equation}
\mbox{\boldmath$D$} \equiv e^{-\mbox{\boldmath${\scriptstyle
\Omega}$}^+} \frac{1}{2} (1+\gamma_5)D
e^{\mbox{\boldmath${\scriptstyle \Omega}$}^+};\qquad
\bar{\mbox{\boldmath$D$}} \equiv e^{\mbox{\boldmath${\scriptstyle
\Omega}$}} \frac{1}{2} (1-\gamma_5) D
e^{-\mbox{\boldmath${\scriptstyle \Omega}$}}.
\end{equation}

\noindent Acting on a field $X$, which is transformed as $X \to
e^{iK} X$, these covariant derivatives are transformed in the same
way. It is also possible to define the covariant derivative with a
Lorentz index

\begin{equation}
\mbox{\boldmath$D$}_\mu \equiv - \frac{i}{4} (C\gamma^\mu)^{ab}
\Big\{\mbox{\boldmath$D$}_a,\bar{\mbox{\boldmath$D$}}_b\Big\},
\end{equation}

\noindent which will have the same property. It is easy to see
that after substitution (\ref{Substitution}) action
(\ref{SYM_Action}) will be

\begin{eqnarray}\label{Background_Action}
&& S = \frac{1}{2e^2}\mbox{tr}\,\mbox{Re}\,\int d^4x\,d^2\theta\,
\mbox{\boldmath$W$}^a \mbox{\boldmath$W$}_a - \frac{1}{64
e^2}\mbox{tr}\,\mbox{Re}\,\int d^4x\,d^4\theta\,\Bigg[16
\Big(e^{-2V}\mbox{\boldmath$D$}^a e^{2V}\Big)
\mbox{\boldmath$W$}_a
+\nonumber\\
&& + \Big(e^{-2V}\mbox{\boldmath$D$}^a e^{2V}\Big)
\bar{\mbox{\boldmath$D$}}^2 \Big(e^{-2V}\mbox{\boldmath$D$}_a
e^{2V}\Big) \Bigg],
\end{eqnarray}

\noindent where

\begin{equation}
\mbox{\boldmath$W$}_a = \frac{1}{32}
e^{\mbox{\boldmath${\scriptstyle \Omega}$}} \bar D (1-\gamma_5) D
\Big(e^{-\mbox{\boldmath${\scriptstyle \Omega}$}}
e^{-\mbox{\boldmath${\scriptstyle \Omega}$}^+} (1+\gamma_5) D_a
e^{\mbox{\boldmath${\scriptstyle \Omega}$}^+}
e^{\mbox{\boldmath${\scriptstyle \Omega}$}} \Big)
e^{-\mbox{\boldmath${\scriptstyle \Omega}$}},
\end{equation}

\noindent and the notation

\begin{eqnarray}\label{Derivatives_Notations}
&& \mbox{\boldmath$D$}^2 \equiv \frac{1}{2} \bar
{\mbox{\boldmath$D$}} (1+\gamma_5)\mbox{\boldmath$D$};\qquad\ \bar
{\mbox{\boldmath$D$}}^2 \equiv \frac{1}{2} \bar
{\mbox{\boldmath$D$}} (1-\gamma_5) \mbox{\boldmath$D$};\nonumber\\
&& \mbox{\boldmath$D$}^a \equiv \Big[\frac{1}{2}\bar
{\mbox{\boldmath$D$}} (1+\gamma_5)\Big]^a;\qquad
\mbox{\boldmath$D$}_a \equiv \Big[\frac{1}{2}(1+\gamma_5)
\mbox{\boldmath$D$}\Big]_a;\nonumber\\
&& \bar {\mbox{\boldmath$D$}}^a \equiv \Big[\frac{1}{2}\bar
{\mbox{\boldmath$D$}} (1 - \gamma_5)\Big]^a;\qquad \bar
{\mbox{\boldmath$D$}}_a \equiv \Big[\frac{1}{2}(1-\gamma_5)
\mbox{\boldmath$D$}\Big]_a
\end{eqnarray}

\noindent is used. Action of the covariant derivatives on the
field $V$ in the adjoint representation is defined by the standard
way.

We note that action (\ref{Background_Action}) is also invariant
under the quantum transformations

\begin{equation}\label{Quantum_Transformations}
e^{2V} \to e^{-\lambda^+} e^{2V} e^{-\lambda};\qquad
\mbox{\boldmath${\Omega}$} \to \mbox{\boldmath${\Omega}$};\qquad
\mbox{\boldmath${\Omega}$}^+ \to \mbox{\boldmath${\Omega}$}^+
\end{equation}

\noindent where $\lambda$ is a background chiral superfield, which
satisfies the condition

\begin{equation}
\bar{\mbox{\boldmath$D$}}\lambda = 0.
\end{equation}

\noindent Such a superfield can be presented as $\lambda =
e^{\mbox{\boldmath${\scriptstyle \Omega}$}} \Lambda
e^{-\mbox{\boldmath${\scriptstyle \Omega}$}} $, where $\Lambda$ is
a usual chiral superfield.

It is convenient to choose a regularization and gauge fixing so
that invariance (\ref{Background_Transformations}) will be
unbroken. First, we fix the gauge by adding the following terms

\begin{equation}\label{Gauge_Fixing}
S_{gf} = - \frac{1}{32 e^2}\,\mbox{tr}\,\int d^4x\,d^4\theta\,
\Bigg(V \mbox{\boldmath$D$}^2 \bar{\mbox{\boldmath$D$}}^2  V + V
\bar {\mbox{\boldmath$D$}}^2 \mbox{\boldmath$D$}^2 V\Bigg)
\end{equation}

\noindent to the action. In this case terms quadratic in the
superfield $V$ will have the simplest form:

\begin{equation}
\frac{1}{2 e^2}\mbox{tr}\,\mbox{Re}\int d^4x\,d^4\theta\, V
\mbox{\boldmath$D$}_\mu^2 V.
\end{equation}

\noindent The corresponding action for the Faddeev--Popov ghosts
$S_{c}$ is written as

\begin{eqnarray}\label{Ghost_Action}
S_{c} = i\,\mbox{tr}\int d^4x\,d^4\theta\,\Bigg\{(\bar c + \bar
c^+) V \Big[(c + c^+) + \mbox{cth}\,V (c-c^+) \Big]\Bigg\}.
\end{eqnarray}

\noindent The superfield $V$ in this expression is decomposed with
respect to the generators of the adjoint representation of the
gauge group, and the fields $c$ and $\bar c$ are the anticommuting
background chiral fields.

Moreover \cite{West}, the quantization procedure also requires
adding the action for the Nielsen--Kallosh ghosts

\begin{eqnarray}\label{B_Action}
S_B = \frac{1}{4e^2}\mbox{tr}\int d^4x\,d^4\theta\,B^+
e^{\mbox{\boldmath${\scriptstyle \Omega}$}^+}
e^{\mbox{\boldmath${\scriptstyle \Omega}$}}\,B,
\end{eqnarray}

\noindent where $B$ is an anticommuting chiral superfield, and the
background field should be decomposed with respect to the
generators of the adjoint representation of the gauge group.
Because the fields $B$ and $B^+$ do not interact with the quantum
gauge field, they contribute only to the one-loop (including
subtraction) diagrams. It is important to note that the factor
$1/e^2$ in action (\ref{B_Action}) is the same as in action for
the gauge fixing terms (\ref{Gauge_Fixing}).

The gauge fixing breaks the invariance of the action under the
quantum gauge transformations (\ref{Quantum_Transformations}), but
there is a remaining invariance under the BRST-transformations.
The BRST-invariance leads to the Slavnov--Taylor identities, which
relate vertex functions of the quantum gauge field and ghosts.
However, all these fields are present only in loops. Later we will
introduce such a regularization that the BRST-invariance is
broken, but background invariance
(\ref{Background_Transformations}) remains unbroken. Then a result
of the calculation is surely gauge invariant (in a sense of the
invariance under the background gauge transformations). A more
complicated question is if it is possible to construct the
renormalized effective action, which satisfies the Slavnov--Taylor
identities. A possibility of using noninvariant regularizations
was investigated in Refs.
\cite{Slavnov1,Slavnov2,Slavnov3,Slavnov4}. According to these
papers, to construct the effective action, satisfying the
Slavnov--Taylor identities, it is necessary to use a special
subtraction scheme, cancelling noninvariant terms in each order of
the perturbation theory. With the background field method this
scheme is slightly simplified, because the background gauge
invariance guarantees, for example, the transversality of the
two-point Green function for the gauge field. Nevertheless, as
earlier, there are additional subtractions in the Green functions,
containing the ghost fields.

However, it is necessary to clear up if using this regularization
affects the result of calculating the Gell-Mann--Low function,
which is investigated in this paper. To answer this question, as a
starting point we will use the following statement: If we fix a
normalization point $\mu\ll\Lambda$ and impose in this point the
boundary condition for the renormalized two-point Green function
$d(p/\mu=1)$, then the two-point Green function is uniquely
determined and does not depend on both a way of renormalization
and a regularization. For example, if two different
regularizations are used, then

\begin{equation}\label{D_Equality}
d_1\Big(\alpha_1(\mu),\frac{p}{\mu}\Big) =
d_2\Big(\alpha_2(\mu),\frac{p}{\mu}\Big),
\end{equation}

\noindent where $\alpha_i(\mu)$ and $d_i$ are the renormalized
coupling constants at the scale $\mu$, and the renormalized
two-point Green functions, obtained in the first and in the second
regularization respectively. Setting $p=\mu$ in Eq.
(\ref{D_Equality}) it is possible to find the dependence
$\alpha_1(\alpha_2)$. Therefore, two different regularizations
differ in a finite renormalization of the coupling constant. We
note that such a renormalization can be gauge dependent and causes
the gauge dependence of the effective action divergent part in a
sufficiently large order of the perturbation theory. However, the
Gell-Mann--Low function, which we will calculate below in this
paper, does not depend on such finite renormalization, because
(setting $x\equiv \ln p/\mu$)

\begin{equation}
\beta_1\Big(d_1(\alpha_1,x)\Big) = \frac{\partial}{\partial x}
d_1(\alpha_1,x) =\frac{\partial}{\partial x} d_2(\alpha_2,x) =
\beta_2\Big(d_2(\alpha_2,x)\Big)=
\beta_2\Big(d_1(\alpha_1,x)\Big).
\end{equation}

\noindent Therefore, the Gell-Mann--Low function does not depend
on a regularization. In particular, a regularization can break the
BRST-invariance, provided the renormalized action is obtained by
subtractions, restoring the Slavnov--Taylor identities. The
Gell-Mann--Low function is gauge independent, because the
dependence of the RHS on the gauge is factorized to the gauge
dependence of the $\beta$-function argument. Therefore, using a
regularization, breaking the BRST-invariance, is possible. In
particular, we will add  the term

\begin{eqnarray}\label{Regularized_Action}
&& S_{\Lambda} = \frac{1}{2 e^2}\mbox{tr}\,\mbox{Re}\int
d^4x\,d^4\theta\, V\frac{(\mbox{\boldmath$D$}_\mu^2)^{n+1}}{
\Lambda^{2n}} V.
\end{eqnarray}

\noindent to action (\ref{Background_Action}).

Proposed way of the regularization and gauge fixing preserves both
invariance under the supersymmetry transformations and the
invariance under transformations
(\ref{Background_Transformations}). As a consequence, the
effective action, calculated with the background field method,
will be invariant under both supersymmetry and gauge
transformations.

We note that the regularization, described here, is different from
a method, proposed in Ref. \cite{West_Paper}. They differ in form
of the term with higher covariant derivatives. In the method,
considered here, it breaks the BRST-invariance, but terms,
quadratic in the quantum superfield $V$, are simpler. This
simplifies calculations in a certain degree, but all typical
features of the higher derivative regularization are the same in
the both cases. However, we should keep in mind that it is
necessary to use a special subtraction scheme, because the higher
derivative term breaks the BRST-invariance. This scheme cancels
noninvariant terms, and ensures that the Slavnov--Taylor
identities are satisfied in each order of the perturbation theory.
It will be discussed in Sec. \ref{Section_Anomaly_Puzzle} in more
details.

Let us construct the generating functional as follows:

\begin{eqnarray}\label{Generating_Functional}
&& Z[J,\mbox{\boldmath$\Omega$}] = \int D\mu\,\exp\Big\{i S + i
S_\Lambda + i S_{gf} + i S_{gh}
+\nonumber\\
&& \qquad\qquad\qquad\qquad\qquad + i \int d^4x\,d^4\theta\,\Big(J
+ J[\mbox{\boldmath$\Omega$}] \Big)
\Big(V'[V,\mbox{\boldmath$\Omega$}] - {\bf V} \Big) \Big\},\qquad
\end{eqnarray}

\noindent where the superfield ${\bf V}$ is given by

\begin{equation}\label{Background Field}
e^{2{\bf V}} \equiv e^{\mbox{\boldmath${\scriptstyle \Omega}$}^+}
e^{\mbox{\boldmath${\scriptstyle \Omega}$}},
\end{equation}

\noindent and $J[\mbox{\boldmath$\Omega$}]$ is a so far undefined
functional. A reason of its introducing will be clear later.
$S_{gf}$ denotes gauge fining terms (\ref{Gauge_Fixing}), and
$S_{gh} = S_c + S_B$ is the corresponding action for the
Faddeev--Popov and Nielsen--Kallosh ghosts. The functional
integration measure is written as

\begin{equation}
D\mu = DV\,D\bar c\,Dc\,DB.
\end{equation}

\noindent We will assume that the coupling constant $e$ is
replaced by the bare coupling constant $e_0$ in all expressions.

In order to understand how generating functional
(\ref{Generating_Functional}) is related with the ordinary
effective action, we perform the substitution $V \to V'$. Then we
obtain

\begin{equation}
Z[J,\mbox{\boldmath$\Omega$}] = \exp\Big\{ -i \int
d^4x\,d^4\theta\,\Big(J + J[\mbox{\boldmath$\Omega$}]\Big) {\bf V}
\Big\} Z_0\Big[J + J[\mbox{\boldmath$\Omega$}],
\mbox{\boldmath$\Omega$}\Big],
\end{equation}

\noindent where

\begin{equation}
Z_0[J,\mbox{\boldmath$\Omega$}] = \int D\mu\,\exp\Big\{i S + i
S_\Lambda + i S_{gf} + i S_{gh} + i \int d^4x\,d^4\theta\,J V
\Big\}.
\end{equation}

\noindent If the dependence of $S$, $S_\Lambda$, $S_{gf}$, and
$S_{gh}$ on the arguments $V$, $\mbox{\boldmath$\Omega$}$, and
$\mbox{\boldmath$\Omega^+$}$ were factorized into the dependence
on the variable $V'$, $Z_0$ would not depend on
$\mbox{\boldmath$\Omega$}$ and $\mbox{\boldmath$\Omega^+$}$ and
would coincide with the ordinary generating functional. This
really takes place for action (\ref{SYM_Action}). However, in the
term with the higher derivatives, in the gauge fixing terms, and
in the ghost Lagrangian such factorization does not occur.
Therefore, $Z_0$ actually differs from the ordinary generating
functional.

Using the functional $Z[J,\mbox{\boldmath$\Omega$},j]$ it is
possible to construct the generating functional for the connected
Green functions

\begin{equation}
W[J,\mbox{\boldmath$\Omega$}] = -i\ln
Z[J,\mbox{\boldmath$\Omega$}] = - \int d^4x\,d^4\theta \Big(J +
J[\mbox{\boldmath$\Omega$}]\Big) {\bf V} + W_0\Big[J +
J[\mbox{\boldmath$\Omega$}],\mbox{\boldmath$\Omega$}\Big].\qquad
\end{equation}

\noindent Also it is possible to construct the corresponding
effective action

\begin{equation}\label{Effective_Action_Definition}
\Gamma[V,\mbox{\boldmath$\Omega$}] = -\int d^4x\,d^4\theta\,\Big(J
{\bf V} + J[\mbox{\boldmath$\Omega$}] {\bf V}\Big) + W_0\Big[J +
J[\mbox{\boldmath$\Omega$}],\mbox{\boldmath$\Omega$}\Big] - \int
d^4x\,d^4\theta\,J V,
\end{equation}

\noindent where the sources should be expressed in terms of fields
using the equation

\begin{eqnarray}
&& V = \frac{\delta}{\delta J} W[J,\mbox{\boldmath$\Omega$}] = -
{\bf V} + \frac{\delta}{\delta J} W_0\Big[J +
J[\mbox{\boldmath$\Omega$}],\mbox{\boldmath$\Omega$}\Big].
\end{eqnarray}

\noindent Substituting this expression into Eq.
(\ref{Effective_Action_Definition}), we write the effective action
as

\begin{eqnarray}
&& \Gamma[V,\mbox{\boldmath$\Omega$}] = W_0\Big[J +
J[\mbox{\boldmath$\Omega$}],\mbox{\boldmath$\Omega$}\Big] - \int
d^4x\,d^4\theta\,\Big(J[\mbox{\boldmath$\Omega$}] {\bf V} + J
\frac{\delta}{\delta J} W_0\Big[J +
J[\mbox{\boldmath$\Omega$}],\mbox{\boldmath$\Omega$}\Big]\Big).
\end{eqnarray}

\noindent Let us now set $V = 0$, so that

\begin{equation}\label{Phi_To_Zero}
{\bf V} = \frac{\delta}{\delta J} W_0\Big[J +
J[\mbox{\boldmath$\Omega$}],\mbox{\boldmath$\Omega$}\Big].
\end{equation}

We also take into account that the invariance under background
gauge transformations (\ref{Background_Transformations})
essentially restricts the form of the effective action. If the
quantum field $V$ in the effective action is set to 0, then the
superfield $K$ will be only in the gauge transformation law of the
fields $\mbox{\boldmath$\Omega$}$ and
$\mbox{\boldmath$\Omega$}^+$, the only invariant combination being
expression (\ref{Background Field}). (It is invariant in a sense
that the corresponding transformation law does not contain the
superfield $K$.) This means that in the final expression for the
effective action we can set

\begin{equation}\label{K_Fixing}
\mbox{\boldmath$\Omega$} = \mbox{\boldmath$\Omega$}^+ = {\bf V}.
\end{equation}

\noindent In this case the effective action is

\begin{eqnarray}\label{Background_Gamma}
&& \Gamma[0,{\bf V}] = W_0\Big[J + J[{\bf V}],{\bf V}\Big] - \int
d^4x\,d^4\theta\,\Big(J + J[{\bf V}]\Big) \frac{\delta}{\delta J}
W_0\Big[J + J[{\bf V}], {\bf V}\Big].
\end{eqnarray}

\noindent Note that this expression does not depend on form of the
functional $J[\mbox{\boldmath$\Omega$}]$. In particular, it can be
chosen to cancel terms linear in the field $V$ in Eq.
(\ref{Generating_Functional}). Such a choice will be very
convenient below.

If the gauge fixing terms, ghosts, and the terms with higher
derivatives depended only on $V'$, expression
(\ref{Background_Gamma}) would coincide with the ordinary
effective action. However, as we already mentioned above, the
dependence on $V$, $\mbox{\boldmath$\Omega$}$, and
$\mbox{\boldmath$\Omega$}^+$ is not factorized into the dependence
on $V'$ with the proposed method of regularization and gauge
fixing. According to Ref. \cite{Kluberg1,Kluberg2}, the invariant
charge (and, therefore, the Gell-Mann--Low function) is gauge
independent, and the dependence of the effective action on gauge
can be eliminated by renormalization of the wave functions of the
gauge field, ghosts, and matter fields. Therefore, for calculating
the Gell-Mann-Low function we may use the background gauge
described above. We note that if this gauge is used, the
renormalization constant of the gauge field $A_\mu$ is 1 due to
the invariance of the action under transformations
(\ref{Background_Transformations}). We note that, as we already
mentioned above,  using a regularization, breaking the
BRST-invariance does not change the Gell-Mann--Low function.

Nevertheless, generating functional (\ref{Generating_Functional})
is not yet completely constructed. The matter is that adding the
term with higher derivatives does not remove divergences from
one-loop diagrams. To regularize them, it is necessary to insert
the Pauli-Villars determinants in the generating functional
\cite{Slavnov_Book}. The Pauli-Villars fields should be introduced
for the quantum gauge field and ghosts (including the
Nielsen--Kallosh ghosts). Constructing them we we will at once use
condition (\ref{K_Fixing}).

So, we insert in the generating functional the factors

\begin{equation}\label{PV_Insersion}
\prod\limits_i \Big(\det PV(V,{\bf V},M_i)\Big)^{c_i},
\end{equation}

\noindent in which the Pauli-Villars determinants are defined by

\begin{equation}\label{PV_Determinants}
\Big(\det PV(V,{\bf V},M)\Big)^{-1} = \int DV_{PV}D\bar
c_{PV}Dc_{PV}DB_{PV}\exp\Big(i S_{PV}\Big),
\end{equation}

\noindent where the action for the Pauli-Villars fields is

\begin{eqnarray}
&& S_{PV}\equiv \mbox{tr}\,\mbox{Re}\int d^4x\,d^4\theta\, V_{PV}
\Big[\frac{1}{2 e_0^2}\mbox{\boldmath$D$}_\mu^2\Big(1 +
\frac{\mbox{\boldmath$D$}_\mu^{2n}}{\Lambda^{2n}}\Big) -
\frac{1}{e_0^2} \mbox{\boldmath$W$}^a \mbox{\boldmath$D$}_a +
\frac{1}{e^2} M_{V}^2\Big]V_{PV} +\nonumber\\
&& + \frac{1}{4}\mbox{tr}\int d^4x\,d^4\theta (\bar c_{PV} + \bar
c_{PV}^+) V \Big[(c_{PV} + c_{PV}^+) + \mbox{cth}\,V
(c_{PV}-c_{PV}^+) \Big] +\nonumber\\
&& + \Bigg(\frac{1}{2}M_c\,\mbox{tr}\int d^4x\,d^2\theta\,\bar
c_{PV}\, c_{PV} + \mbox{h.c.}\Bigg) +\frac{1}{4e_0^2}\mbox{tr}\int
d^4x\,d^4\theta\,B_{PV}^+ e^{2{\bf V}} B_{PV} +\nonumber\\
&& + \mbox{tr}\Bigg(\frac{1}{2e^2}\int d^4x\,d^2\theta\,M_B
B_{PV}^2 + \mbox{h.c.}\Bigg).
\end{eqnarray}

\noindent The Grassmanian parity of the Pauli--Villars fields is
opposite to the Grassmanian parity of usual fields, corresponding
to them. The coefficients $c_i$ in Eq. (\ref{PV_Insersion})
satisfy conditions

\begin{equation}
\sum\limits_i c_i = 1;\qquad \sum\limits_i c_i M_i^2 = 0.
\end{equation}

\noindent Below, we assume that $M_i = a_i\Lambda$, where $a_i$
are some constants. Inserting the Pauli-Villars determinants
allows cancelling the remaining divergences in all one-loop
diagrams, including diagrams containing counterterm insertions.
(This is guaranteed because the masses of the gauge field and
Nielsen--Kallosh ghosts are multiplied by the renormalized
coupling constants, and the other terms are multiplied by the bare
ones. This will be discussed later in more details.)

\section{Two-loop calculation}
\label{Section_2Loop} \hspace{\parindent}

The one-loop $\beta$-function, calculated with the background
field method, is well-known \cite{West}. Using the higher
covariant derivative regularization does not essentially change
the calculation, and its result \cite{PhysLett}. Let us mention
the typical features. The quantum superfield $V$ does not
contribute to the one-loop diagrams, because in the corresponding
diagrams the number of the spinor derivatives $D$, acting on
propagators, is less than 4. Really, a result of calculating any
two-point diagram is proportional to

\begin{equation}
\delta^8_{xy}\,\hat P \delta^8_{xy},
\end{equation}

\noindent where $x$ and $y$ are the points, to which the external
lines are attached. The result is not 0 only if the operator $\hat
P$ contains 4 spinor derivatives. However, two vertexes can
contain no more than 2 spinor derivatives, and propagators of the
gauge field do not contain spinor derivatives at all. Therefore,
all one-loop two-point diagrams are automatically 0. The one-loop
diagrams with the Pauli--Villars fields, corresponding to the
gauge field, are 0 due to the same reason. Because the higher
derivatives do not change a number of spinor derivatives in
vertexes, the one-loop contribution of the quantum field is also 0
in the regularized theory.

Therefore, the one-loop two-point Green-function of the gauge
field is completely determined by contributions of the
Faddeev--Popov and Nielsen--Kallosh ghosts. With the
regularization and gauge fixing, described above, the ghost
Lagrangians do not depend on the presence of higher derivative
terms. Due to anticommuting, the contributions of each ghost
fields have opposite sign in comparison with the contribution of
the chiral scalar superfield in the adjoint representation of the
gauge group. Therefore, in the one-loop approximation the
Gell-Mann--Low function is

\begin{equation}
\beta(\alpha) = - \frac{3 C_2 \alpha^2}{2\pi} + O(\alpha^3).
\end{equation}

\begin{figure}[h]
\includegraphics{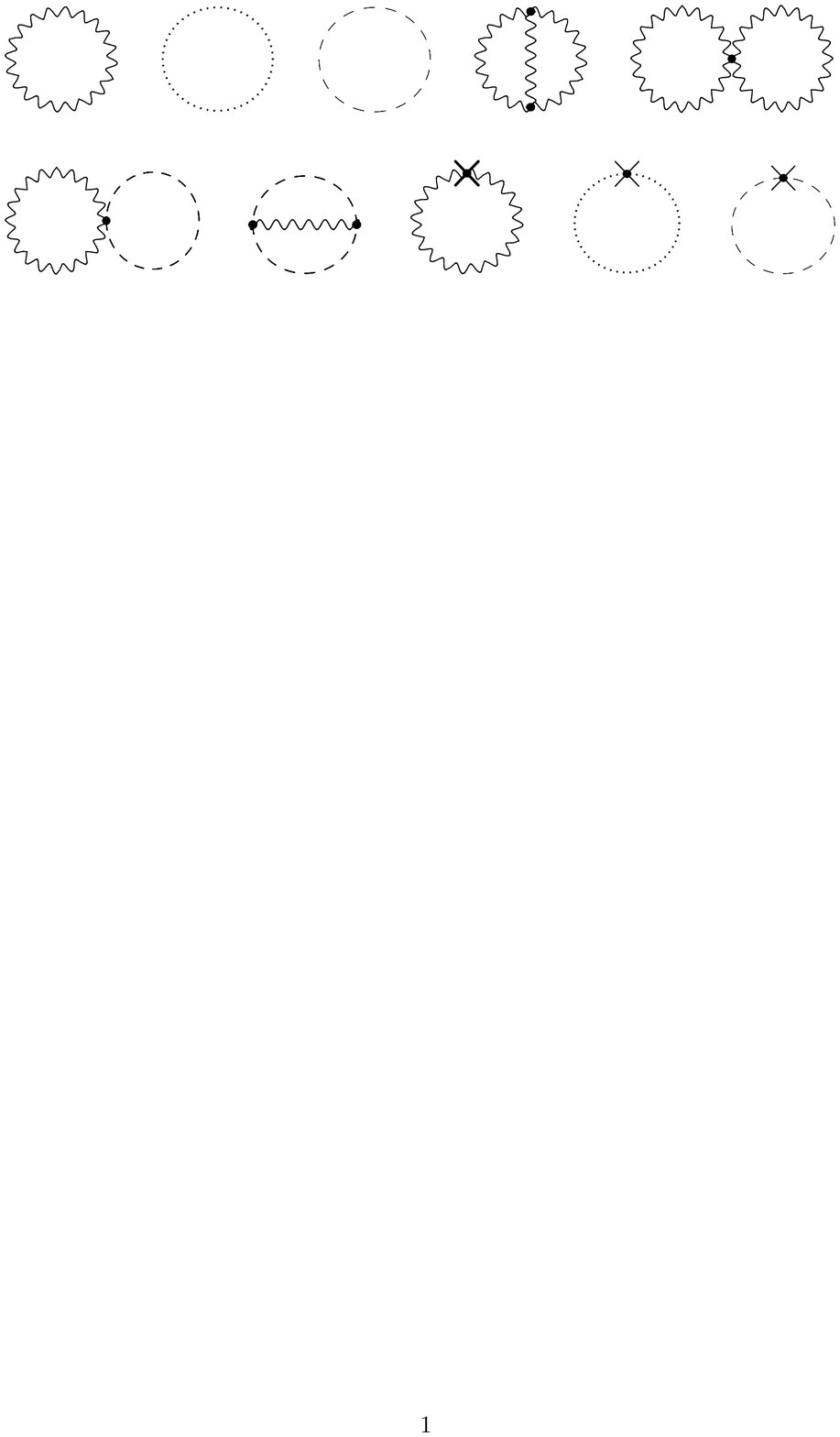}
\caption{Diagrams, contributing to the two-loop $\beta$-function
of the $N=1$ supersymmetric Yang--Mills theory.}
\label{Figure_Diagrams}
\end{figure}

The effective action in the two-loop approximation is calculated
by the standard way. It is contributed by diagrams, schematically
presented in Fig. \ref{Figure_Diagrams}. Usual diagrams are
obtained by attaching to them two external lines of the background
gauge field by all possible ways. In Fig. \ref{Figure_Diagrams} a
propagator of the quantum field $V$ is denoted by a wavy line, a
propagator of the Faddeev--Popov ghosts by dashes, and a
propagator of the Nielsen--Kallosh ghosts by dots. (We note that
they contribute only in the one-loop approximation, because the
Nielsen--Kallosh ghosts interact only with the background field.)

With the higher derivative regularization the propagator of the
quantum field is

\begin{equation}
\frac{1}{q^2 (1+q^{2n}/\Lambda^{2n})}
\end{equation}

\noindent (in the Euclidean space after the Weak rotation).
Feynman rules for vertexes, containing two lines of the quantum
field $V$, are also changed. In particular, a vertex with a single
line of the background superfield ${\bf V}$, which has the
momentum $p$, (it is denoted by a bold wavy line) is

\begin{equation}
\hspace*{-11cm}
\includegraphics[scale=0.5]{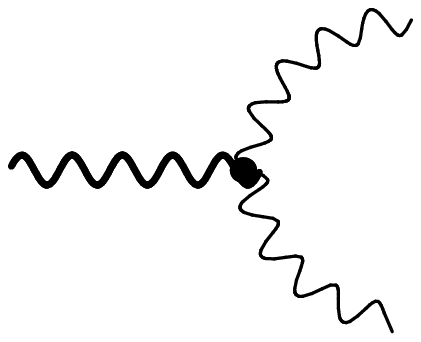}
\begin{picture}(-2,-0.7)(-2,-0.7)
\put(-2,0){${\displaystyle \sim \frac{1}{4}(2k+p)_\mu \bar D
\gamma^\mu\gamma_5 D {\bf V}
\Bigg(1+\frac{(k+p)^{2n+2}-k^{2n+2}}{\Lambda^{2n}\Big((k+p)^2-k^2\Big)}
\Bigg),}$}
\end{picture}
\end{equation}

\noindent and a vertex with two lines of the background superfield
${\bf V}$, which have momentums $p$ and $-p$ is

\begin{eqnarray}
&& \hspace*{-0.7cm}\includegraphics[scale=0.6]{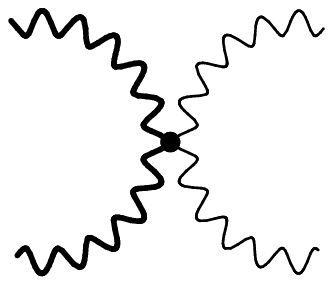}
\begin{picture}(-2,-0.7)(-2,-0.7)
\put(-2,0){${\displaystyle \sim \Big(4 {\bf V}
\partial^2\Pi_{1/2}{\bf V} +p^2{\bf
V}^2\Big)\Big(1+(n+1)\frac{k^{2n}}{\Lambda^{2n}}\Big)
+\frac{1}{\Lambda^{2n}}\Big((2k+p)^2{\bf V}
\partial^2\Pi_{1/2}{\bf V}+}$}
\end{picture}\nonumber\\
&&  + {\bf V}^2((k+p)^2-k^2)^2 \Big)
\Bigg(\frac{(k+p)^{2n+2}-k^{2n+2}}{((k+p)^2-k^2)^2} -
\frac{(n+1)k^{2n}}{(k+p)^2-k^2}\Bigg) -4{\bf
V}\partial^2\Pi_{1/2}{\bf V}.\qquad
\end{eqnarray}

According to the performed calculations, the two-loop contribution
of the Faddeev--Popov ghosts to the Gell-Mann--Low function is 0
that agrees, for example, with Ref. \cite{Grisaru}. (The
integrals, defining the two-point Green function, appeared to be
some finite constants for the ghosts.)

As we already mentioned, the total two-loop contribution of the
two-point diagrams to the effective action can be presented in
form (\ref{D_Definition}) due to the Slavnov--Taylor identity. To
find the function $d^{-1}$ up to an unessential constant, we
differentiate it with respect to $\ln\Lambda$, and then set the
external momentum $p$ to 0. (This is possible due to using the
higher covariant derivative regularization.) Later we will see
that the result is a some finite constant $d_2$:

\begin{equation}
\frac{d}{d\ln\Lambda} d^{-1}(\alpha,\Lambda/p)\Bigg|_{p=0} = d_2.
\end{equation}

\noindent Therefore, the function $d^{-1}$ depends on the momentum
logarithmically

\begin{equation}
d^{-1}(\alpha,\Lambda/p) = d_2 \ln\frac{\Lambda}{p} +\mbox{const}.
\end{equation}

Calculating explicitly two-loop diagrams, presented in Fig.
\ref{Figure_Diagrams} (so far without diagrams with counterterm
insertions), differentiating the result with respect to
$\ln\Lambda$, and, then, setting $p=0$, we obtain (in the
Euclidean space, after the Weak rotation)

\begin{eqnarray}\label{D2}
&& d_2 = 8\pi\cdot 6\pi\,\alpha_0\,\frac{d}{d\ln\Lambda} \int
\frac{d^4q}{(2\pi)^4} \Big(q^2(1+q^{2n}/\Lambda^{2n})\Big)^{-1}
\int \frac{d^4k}{(2\pi)^4} \frac{1}{k^2}
\frac{d}{dk^2}\Bigg\{\Big((q+k)^2\times\nonumber\\
&& \times (1+(q+k)^{2n}/\Lambda^{2n})\Big)^{-1} \Bigg[2(n+1)
\Big(1+k^{2n}/\Lambda^{2n}\Big)^{-1} - 2n
\Big(1+k^{2n}/\Lambda^{2n}\Big)^{-2} \Bigg]\Bigg\}.
\end{eqnarray}

\noindent It is important to note that taking a limit $p\to 0$ is
rather nontrivial, because the final result can contain infrared
divergent terms, proportional to $p$ or $p^2$, or terms,
proportional to $p$, but giving a finite contribution to $d_2$.
However, the calculation shows that all such terms are cancelled.
Moreover, the sum of diagrams appeared to be a total derivative
with respect to the module of the loop momentum, so that the
integral over $d^4k$, which is contained in Eq. (\ref{D2}), can be
easily calculated. Really, in the four-dimensional spherical
coordinates

\begin{equation}
\int \frac{d^4k}{(2\pi)^4}\frac{1}{k^2} \frac{d}{dk^2} f(k^2) =
\frac{1}{16\pi^2} \Big(f(k^2=\infty) - f(k^2=0)\Big).
\end{equation}

\noindent All substitutions at the upper limit are 0 due to the
higher derivative regularization, and only the substitution at the
lower limit is nonzero. Using equations, presented above, we
obtain

\begin{equation}
d_2 = - 6\alpha_0 \frac{d}{d\ln\Lambda} \int \frac{d^4q}{(2\pi)^4}
\Big(q^2(1+q^{2n}/\Lambda^{2n})\Big)^{-2}.
\end{equation}

\noindent This integral can be also easily calculated in the
four-dimensional spherical coordinates:

\begin{equation}
d_2 = \frac{12\alpha_0}{\pi} \int \frac{d^4q}{(2\pi)^4}
\frac{1}{q^4} q^2 \frac{d}{dq^2}(1+q^{2n}/\Lambda^{2n})^{-2} =
\frac{3\alpha_0}{4\pi^3}(1+q^{2n}/\Lambda^{2n})^{-2}\Bigg|_{0}^\infty
= - \frac{3\alpha_0}{4\pi^2}.
\end{equation}

\noindent (We note that the result does not depend on the
regularization parameter $n$.) Therefore, in the two-loop
approximation

\begin{equation}\label{Two_Loop_D}
d^{-1}(\alpha_0,\Lambda/p) = \frac{1}{\alpha_0} - \frac{3
C_2}{2\pi} \ln\frac{\Lambda}{p} - \frac{3 \alpha_0
C_2^2}{(2\pi)^2}\ln\frac{\Lambda}{p} + O(\alpha_0^2).
\end{equation}

\noindent Therefore, the Gell-Mann--Low function, defined by Eq.
(\ref{Gell-Mann-Low_Definition}), in the two-loop approximation is

\begin{equation}
\beta(\alpha) = - \frac{3 C_2 \alpha^2}{2\pi} - \frac{3\alpha^3
C_2^2}{(2\pi)^2} + O(\alpha^4),
\end{equation}

\noindent and coincides with the expansion of the exact NSVZ
$\beta$-function in the considered order. We note that this result
does not depend on a possible finite constant in Eq.
(\ref{Two_Loop_D}).

\section{Calculating diagrams with counterterms insertions}
\hspace{\parindent} \label{Section_Anomaly_Puzzle}

For calculating quantum corrections it is also necessary to take
into account diagrams with counterterm insertions. Usually, adding
counterterms is equivalent to splitting the bare coupling constant
into the renormalized coupling constant and some infinite
additional term. However, using noninvariant regularizations (and,
in particular, the regularization, breaking the BRST-invariance,
which is used here), it is also necessary to add counterterms,
restoring the Slavnov--Taylor identities \cite{Slavnov1,Slavnov2}
in each order of the perturbation theory. However, in the
considered case the situation is slightly simplified. Really, the
one-loop two-point Green function for the Faddeev--Popov ghosts is
finite and does not depend on regularization. Interaction of
ghosts with the background field is fixed by the background gauge
invariance, which is unbroken in the considered regularization.
Therefore, additional counterterms do not contribute to
subtraction diagrams, containing a loop of the Faddeev--Popov
ghosts, in the two-loop approximation. Moreover, terms with the
Faddeev--Popov ghosts do not evidently depend on whether bare or
renormalized coupling constant is in the gauge fixing action.
Therefore, their contributions do not also depend on a way of
splitting the bare coupling constant into the renormalized one and
counterterms.

Quantizing the theory we also write the bare coupling constant
$e_0$ in the gauge fixing terms. Therefore, a part of the action,
quadratic in the quantum field, is written as

\begin{equation}
\frac{1}{2 e_0^2}\mbox{tr}\,\mbox{Re}\int d^4x\,d^4\theta\, V
\Big[\mbox{\boldmath$D$}_\mu^2\Big(1 +
\frac{\mbox{\boldmath$D$}_\mu^{2n}}{\Lambda^{2n}}\Big) + 2 W^a
\mbox{\boldmath$D$}_a\Big]V.
\end{equation}

\noindent Breaking the invariance under the BRST-transformations
can lead to the necessity of adding counterterms proportional to

\begin{equation}
\mbox{tr}\int d^4x\,d^4\theta V \mbox{\boldmath$D$}_\mu^2 V.
\end{equation}

\noindent (If the background field is 0, this follows from Refs.
\cite{Slavnov3,Slavnov4}. Terms, containing the background field,
can be restored due to the background gauge invariance.) But this
means that all one-loop diagrams, {\it including diagrams with
insertions both of the counterterms, appearing due to the
renormalization of the coupling constant, and of the additional
counterterms}, with a loop of the quantum field $V$, are 0,
because they can contain no more than 2 spinor derivatives.

At last, let us consider diagrams, containing a loop of the
Nielsen--Kallosh ghosts. Because the Nielsen--Kallosh ghosts exist
only in the one-loop approximation, there are no additional
counterterms, caused by the noninvariance of the regularization
under the BRST-transformations, in these diagrams. However, the
contribution of the counterterm diagrams is essential due to the
renormalization of the coupling constant. Really, the coefficient
in the action for the Nielsen--Kallosh ghosts should be the same
as in the gauge fixing terms. Therefore, it must contain the bare
coupling constant:

\begin{equation}
\frac{1}{4 e_0^2} \mbox{tr}\int d^8x\,B^+ e^{2{\bf V}} B.
\end{equation}

\noindent To regularize diagrams with counterterm insertions and a
loop of Nielsen--Kallosh ghosts, the action for the corresponding
Pauli--Villars fields should be written as

\begin{equation}\label{PV_Action}
\mbox{tr}\int d^4x\,\Big(\frac{1}{4 e_0^2}\int d^4\theta\,
B_{PV}^+ e^{2{\bf V}} B_{PV} + \frac{M_B}{2e^2} \int d^2\theta
B_{PV}^2 + \frac{M_B}{2e^2} \int d^2\theta (B_{PV}^+)^2\Big),
\end{equation}

\noindent where $M_B$ is proportional to the regularization
parameter $\Lambda$. Really, let us present a bare coupling
constant as

\begin{equation}
\frac{1}{e_0^2} = \frac{1}{e^2} Z_3,
\end{equation}

\noindent where $e$ is the renormalized coupling constant, and
$Z_3$ is the renormalization constant. Then, expanding the
Pauli--Villars determinant for the Nielsen--Kallosh ghosts in
powers of $Z_3-1$, we obtain terms, regularizing diagrams with
insertions of counterterms.

However, due to inserting this determinant the generating
functional starts to depend on the normalization point at the
fixed bare coupling constant $e_0$, because the renormalized
coupling constant $e$ depends on $\mu$.

In the Abelian case calculating divergences for the action,
similar to (\ref{PV_Action}), was made, for example, in Ref.
\cite{hep}. In the considered case it is also necessary to take
into account a factor $-C_2/2$, which appears because the
Nielsen--Kallosh ghosts are in the adjoint representation of a
gauge group and anticommute. (There is only one matter superfield
now, instead of 2 matter superfields in the Abelian case.)
Moreover, the renormalization constant of the matter field $Z$
should be substituted for the constant $Z_3$. Taking into account
this comments, the result of Ref. \cite{hep} can be formulated as
follows. Contribution of the counterterm diagrams for the
Nielsen--Kallosh ghosts to $1/d$ can be written as

\begin{equation}
\frac{C_2}{2\pi} \ln Z_3.
\end{equation}

To find this contribution in the two-loop approximation we note
that after the one-loop renormalization the renormalization
constant will be

\begin{equation}
Z_3 = 1 + \frac{3C_2 \alpha}{2\pi}\ln\frac{\Lambda}{\mu} +
O(\alpha^2).
\end{equation}

\noindent Therefore, the contribution of diagrams with counterterm
insertions in the two-loop approximation is written as

\begin{equation}
\frac{3 \alpha C_2^2}{(2\pi)^2} \ln\frac{\Lambda}{\mu}.
\end{equation}

\noindent This contributions exactly cancels the two-loop
divergence so that after the one-loop renormalization

\begin{equation}
d^{-1}(\alpha,\mu/p) = \frac{1}{\alpha} - \frac{3 C_2}{2\pi}
\ln\frac{\mu}{p} - \frac{3 \alpha C_2^2}{(2\pi)^2}\ln\frac{\mu}{p}
+ O(\alpha^2).
\end{equation}

For an arbitrary order of the perturbation theory it is reasonable
to propose that the two-point Green function of the gauge field is
given by

\begin{equation}\label{Alpha}
\frac{1}{d(\alpha,\mu/p)} = \frac{1}{\alpha_0} - \frac{1}{2\pi}
C_2 \ln d(\alpha_0,\Lambda/p) + \frac{1}{2\pi} C_2 \ln
Z_3(\alpha,\Lambda/\mu) - \frac{3}{2\pi} C_2 \ln\frac{\Lambda}{p}.
\end{equation}

\noindent Really, it is easy to see that the exact NSVZ
$\beta$-function is obtained by differentiating this equality with
respect to $\ln p$, and the term, proportional to $\ln Z_3$ is
obtained from contributions of diagrams with counterterm
insertions. In the two-loop approximation this equation agrees
with (\ref{Two_Loop_D}) if the contribution of diagrams with
counterterm insertions is taken into account.

If Eq. (\ref{Alpha}) is true, then divergences exist only in the
one-loop approximation. Really, because

\begin{equation}
\frac{1}{d(\alpha,\mu/p)} =
\frac{1}{d(\alpha_0(\alpha,\Lambda/\mu),\Lambda/p)}
Z_3(\alpha,\Lambda/\mu)
\end{equation}

\noindent is finite, it is necessary to cancel only the one-loop
divergence. For this purpose the bare coupling constant is
presented as

\begin{equation}
\frac{1}{\alpha_0} = \frac{1}{\alpha} + \frac{3}{2\pi} C_2
\ln\frac{\Lambda}{\mu}.
\end{equation}

\noindent We note that presence of divergences only in the
one-loop approximation in this case does not mean that the
physical $\beta$-function has only the one-loop contribution.
Really, the physical $\beta$-function is a derivative of the
two-point Green function with respect to the logarithm of the
momentum if proper boundary conditions are imposed. Such function,
as we already saw, has corrections in all loops. A relation
between the divergences and the physical $\beta$-function is
broken due to the way of the regularization of diagrams with the
insertions of counterterms, which leads to the dependence of the
generating functional on the normalization point at the fixed bare
coupling constant \cite{HD_And_DRED}.

So, if Eq. (\ref{Alpha}) is valid, the Gell-Mann--Low function
coincides with the exact NSVZ $\beta$-function, and divergences in
the two-point Green function exist only in the one-loop
approximation.


\section{Conclusion}
\label{Section_Conclusion}
\hspace{\parindent}

Investigation, made in this paper, shows that the higher covariant
derivative regularization can be easily applied for calculating
quantum corrections in the supersymmetric Yang--Mills theory. Its
using allows differentiating with respect to the regularization
parameter $\Lambda$ and setting the external momentum to 0. As a
result, it is possible to find the Gell-Mann--Low function, which
in the considered approximation coincides with the expansion of
the exact NSVZ $\beta$-function. (We note, once again, that the
Gell-Mann--Low function does not depend on the choice of the
renormalization scheme.) A very interesting feature of using the
higher covariant derivative regularization in supersymmetric
theories is a possibility of calculating all integrals
analytically, because their sum is reduced to a total derivative.
Exactly the same feature was noted in the Abelian case
\cite{ThreeLoop}.

With the higher derivative regularization divergences in the
two-point Green function appeared to exist only in the one-loop
approximation. (However, the divergent part of the two-point Green
function is not a physical quantity. A physical quantity is the
Gell-Mann--Low function, which are contributed by all orders of
the perturbation theory.) The obtained result to a considerable
extent confirms conclusions of Ref. \cite{SV}, where the authors
proposed that the Wilson action $S_W$ was renormalized only at the
one-loop, and the effective action $\Gamma$ had corrections in all
loops. In our case the usual renormalized action plays a role of
$S_W$. As for the electrodynamics, the Gell-Mann--Low function
does not coincide with the function $b(\alpha)$, defined by the
divergent part of the effective action, due to the rescaling
anomaly \cite{Arkani}, which leads to the dependence of the
standardly defined generating functional on the normalization
point.

We note that using the higher covariant derivative regularization
can possibly allow deriving the expression for the exact NSVZ
$\beta$-function by the straightforward summation of Feynman
diagrams exactly to  all orders of the perturbation theory,
similar to the case of the supersymmetric electrodynamics. Now
this work is in progress.

\bigskip
\bigskip

\noindent {\Large\bf Acknowledgments.}

\bigskip

\noindent This paper was partially supported by the Russian
Foundation for Basic Research (Grant No. 05-01-00541).


\end{document}